\documentstyle[12pt,a4wide]{article}
\def\As{/\kern-2.6mm A}
\def\bea{\begin{eqnarray}}
\def\beq{\begin{equation}}
\def\Cc{{\cal C}}
\def\de{\delta}
\def\Ds{/\kern-2.6mm D}
\def\eea{\end{eqnarray}}
\def\eeq{\end{equation}}
\def\l{\langle}
\def\Lam{\Lambda_{\rm QCD}}
\def\LH{    {\cal L}^{ {\rm h}   }  }
\def\LPsi{    {\cal L}^\Psi  }
\def\Lhqet{ {\cal L}^{ {\rm HQET}}  }
\def\Lqcd{{\cal L}^{{\rm QCD}}_{{\rm light}}}
\def\LQCD{{\cal L}^{{\rm QCD}}}
\def\MSb{$\overline{{\rm MS}}$}
\def\nn{\nonumber}
\def\NPB#1 #2 #3 {Nucl.~Phys.~{\bf B#1}
(19#2)~#3}
\def\p{\partial}
\def\PLB#1 #2 #3 {Phys.~Lett.~{\bf B#1}
(19#2)~#3}
\def\ps{/\kern-2.3mm p}
\def\Pv{\Sh}
\def\r{\rangle}
\def\Sh{S^\pm_v(p)}
\def\tr{{\rm tr}\,}
\def\vs{/\kern-2.3mm v}
\def\ZF{Z_2}
\def\ZH{Z_h}
\def\ZPsi{Z_\Psi}
\def\Zt{\widetilde{Z}}
\def\ZYM{Z_3}
\begin{document}
\begin{titlepage}
\renewcommand{\thefootnote}{\fnsymbol{footnote}}
\begin{flushright}
UAB-FT-314 \\
hep-ph/9305297
\end{flushright}
\vskip0.4cm
\begin{center}
\renewcommand{\baselinestretch}{1.4}
\boldmath
{\large\bf RENORMALIZABILITY OF THE HEAVY QUARK EFFECTIVE THEORY}
\unboldmath
\vskip1cm
\renewcommand{\baselinestretch}{1.1}
\newlength{\estrecho}
\setlength{\estrecho}{0.15\textwidth}
\addtolength{\textwidth}{-\estrecho}
{\sc E. Bagan\footnote{Alexander von Humboldt fellow.}
and P. Gosdzinsky}\\[1cm]
Grup de F\'\i sica Te\`orica,\\
Departament de F\'\i sica\\
and \\
Institut de F\'\i sica d'Altes Energies,\\[0.5cm]
Universitat Aut\`onoma de Barcelona,\\
E-08193 Bellaterra (Barcelona), Spain.\\[6cm]
\end{center}
\begin{abstract}
We show that the Heavy Quark Effective Theory is renormalizable
perturbatively. We also show
that there exist renormalization schemes in which the infinite
quark mass limit of any QCD Green function is exactly given by the
corresponding Green function of the Heavy Quark Effective Theory.
All this is accomplished while preserving BRS invariance.
\end{abstract}
\renewcommand{\thefootnote}{\arabic{footnote}}
\setcounter{footnote}{0}
\end{titlepage}

                     \newpage

\section{\bf Introduction}
\label{S-in}

Over the last few years there has been an enormous interest in the so called
Heavy Quark Effective Theory (HQET)~\cite{general1}.
Hadrons made out of a heavy quark,
such as $b$ or $c$, and
either a light anti-quark (heavy-light meson) or two
light quarks (heavy-light
baryons) can be conveniently analyzed within the framework of the HQET.
It exploits two symmetries which are not
apparent in the standard QCD lagrangian
when quark masses are small ($\approx\Lam$): the {\em spin} and
{\em flavour}
symmetries. They come down to the statement that
the dynamics of a heavy-light hadron is heavy-flavour
and heavy-quark spin independent.
These two symmetries can be combined into a larger one:
the {\em Isgur-Wise} symmetry~\cite{I&W}.

The phenomenological implications of the Isgur-Wise symmetry
have been extensively discussed in the recent
literature~\cite{general}.
In this letter we address the field theoretical questions of
whether the HQET is renormalizable or not and whether renormalization can be
accomplished or not while preserving BRS invariance.
In applications of the HQET, renormalizability is always assumed and
even some renormalization constants have been computed under this
assumption. However, as far as we know, a formal proof is still missing.
These are not just
academic questions since loop calculations are required
in order to obtain the scaling (or large heavy-quark
mass behaviour) of many phenomenologically relevant quantities, e.g.
leptonic and semileptonic decay constants and form factors.
Although the asymptotic behaviour of these and other parameters could
in principle be obtained from standard QCD~\cite{shi-nov},
the calculations are far more involved than the corresponding ones in the HQET
and, to the best of our knowledge, no attempt has been made beyond
one loop.
This raises another important question: Does the HQET really provide
the right (QCD based) large quark mass result
beyond tree-level? We come back to this point below.

Our first goal is to prove that the
HQET is renormalizable perturbatively to any order. What has to be shown is
that it is possible to rescale the fields and coupling constants so that all
Green functions involving only elementary fields are free of ultraviolet
divergences without spoiling BRS invariance.
(In fact, we would only need a
proof for S-matrix amplitudes. However, these will be finite if Green
functions have no divergences.) Thus, we choose a
regularization scheme which preserves BRS invariance, e.g.
dimensional regularization.

Throughout this letter we will assume that Weinberg's convergence
theorem~\cite{W-T}
is satisfied even though Lorentz invariance is broken by
the heavy quark propagator, $1/p\cdot v$. More explicitly, we will assume
that if a 1PI graph $G$ has superficial degree of divergence $\de(G)$, given by
naive
power counting, then its overall
UV divergence is a polynomial in the external momentum
of at most degree $\de(G)$. We shall not attempt to prove this statement here,
but rather note that the heavy quark propagator $1/p\cdot v$ is very
similar to the principal-value prescribed pole $1/p\cdot n$ of axial gauges
for which one can argue that Weinberg's theorem holds~\cite{kummer}.
Furthermore, a
large variety of
one- and two-loop
calculations~\cite{ko-ku} give support to this assumption.

Our second goal is to show that there exist renormalization schemes
in both
QCD and the HQET consistent with BRS invariance and such that any QCD
Green
function agrees with its
HQET counterpart
up to order $O(1/M)$. We shall refer to this as the {\em matching}.
For instance, for the quark propagator one has
\beq
\l \Psi \bar\Psi\r|_{\pm Mv+p}=\l h^\pm_v \bar h^\pm_v
\r|_p+O\left({1\over M}\right),
\label{e1-a}
\eeq
where $\Psi$ ($M$) is the field (mass) of a heavy
quark, i.e.~$M\gg\Lam$,
and $h^+_v$ ($h^-_v$) is the effective
field in the HQET corresponding to the particle (antiparticle)
part of $\Psi$. In this letter we use the shorthand notation
$\l\cdots\r=\l0|{\bf T}\cdots|0\r$.
Note that on the HQET side of equations such as~(\ref{e1-a}),
the momentum, $p$, of a heavy quark (antiquark) with velocity $v$ is actually
its virtuality or ``off-shellness'', defined by subtracting $Mv$ ($-Mv$)
from
the real momentum used on the QCD side.
Assuming renormalizability of the HQET,
Grinstein~\cite{g1} has shown that
QCD and HQET Green functions match
provided one chooses the appropriate counterterms,
which turn out to depend logarithmically on $M$. Unfortunately,
it is not clear if his approach preserves BRS invariance.
We shall prove the following statement: By choosing our
renormalization scheme in such a way that the heavy quark two-point
functions in QCD and in the HQET agree up to order $O(1/M)$, BRS invariance
ensures also the matching of the heavy quark-gluon vertex and, in turn,
of any other Green function. Hence, our approach
explicitly preserves the
BRS invariance of the theory.

In applications, one usually works in the \MSb\ (or MS) scheme. From
the preceding
paragraph it follows that there must exist a coefficient $C(\log M)$ such that,
for example,
$\l \Psi \bar\Psi\r|_{\pm Mv+p}^{\overline{\rm MS}}=
C(\log M)\l h^\pm_v \bar h^\pm_v \r|_p^{\overline{\rm MS}}
+O(1/M)$. Similar equations hold for any pair of
Green functions. The coefficient
$C^{1/2}(\log M)$ is just the {\em finite} wave
function renormalization constant of the
heavy quark that relates
the \MSb\ scheme with the above-mentioned ones.
It is important because it provides the scaling properties of the QCD
Green functions
at large $M$.

This letter is organized as follows:
In sec.~\ref{sec1b} we introduce the HQET lagrangian and discuss the BRS
symmetry. In sec.~\ref{sec2} we briefly review the proof of
renormalizability of QCD. We will follow very closely the proof given by
Collins \cite{c}. In sec.~\ref{sec3} we will show that the HQET is
renormalizable. In
sec.~\ref{sec4}, we prove that matching of QCD and the HQET can be achieved
while maintaining BRS invariance.
Finally, sec.~\ref{sec5} will be devoted to
comments and conclusions.

\section{The HQET Lagrangian and BRS invariance}
\label{sec1b}

In terms of renormalized fields the lagrangian of the HQET,
$\Lhqet$,
has the form:
\beq
\Lhqet=\LH+\Lqcd,
\label{e-a}
\eeq
where
\bea
\LH&=&\ZH \bar h^+_v(x) i\, v\cdot D h^+_v(x)-
\ZH \bar h^-_v(x) i\, v\cdot D h^-_v(x);\label{e-b}\\
\Lqcd&=&-{1\over 4} \ZYM (G_{\mu \nu}^a)^2
+\ZF\bar\psi (i \Ds -m_0)\psi -{1\over 2\xi}(\partial\cdot A^a)^2+
\Zt{\partial}_\mu \bar c^a D^\mu c^a .
\label{e-c}
\eea
For simplicity, only one heavy and one light quark
flavour
will be considered.
Here,
\beq
G_{\mu \nu}^a=\partial_\mu A_\nu^a - \partial_\nu A_\mu^a -{X\over \Zt }
g c_{abc}A_\mu^b A_\nu^c
\label{e-d}
\eeq
is the (gluon) field-strength tensor. The covariant derivatives in the
defining and adjoint representations of $SU(N)$ are respectively:
\bea
D_\mu \chi&=&(\partial_\mu +ig{X\over \Zt } t^a A_\mu^a)\chi;
\qquad \chi=\psi,h^\pm_v;
\label{e-e}
\\
D_\mu c_a&=&\partial_\mu c^a  +{X\over \Zt }g c_{abc}c^b A_\mu^c .
\label{e-f}
\eea
In eqs.(\ref{e-b}-\ref{e-f}), $g$ is the coupling constant,
$\ZYM$, $\Zt$, $\ZF$ and $\ZH$ are the wave function
renormalization constants of the $A^a_\mu$ (gluon), $c^a$ (ghost),
$\psi$ (light quark) and $h^\pm_v$ (effective
heavy quark/antiquark) fields, whereas
$X$ renormalizes the ghost-gluon vertex. Finally, $m_0$ is the
bare (light) quark  mass, $t^a$ are the generators of $SU(N)$ in its
defining representation satisfying
$\tr (t^a t^b)={1\over 2}\delta^{ab}$ and $c_{abc}$ are the structure
constants defined through $[t^a,t^b]=
ic_{abc}t^c$.

The global symmetries of $\Lhqet$, particularly the Isgur-Wise symmetry,
have been discussed (see for instance~\cite{gen-geor}) and
exploited extensively
in the literature. It has also been shown from different
approaches~\cite{georgi}
that {\em at tree-level} the lagrangian
$\Lhqet$ can be obtained from the full
QCD lagrangian $\LQCD=\LPsi+\Lqcd$, where
\beq
\LPsi=\ZPsi\bar\Psi (i \Ds -M_0)\Psi
\label{e-_g}
\eeq
describes an ordinary
quark field $\Psi$ with mass $M\gg \Lam $ coupled to gluons
through the covariant derivative
$D_\mu$ defined as in~(\ref{e-e}).
In the \MSb\ scheme one has $\ZPsi=\ZF$, however, if a mass dependent
scheme is chosen, as we shall do to prove the matching in section~\ref{sec4},
two independent wave function renormalization constants are required.

The lagrangian (\ref{e-a}) is invariant under the
(renormalized) BRS transformations\break
$\delta_{BRS}\varphi=\delta_R\varphi\;\delta\lambda_R$
where $\varphi$ is any basic field, $\de\lambda_R$ is a Grassmann number
and $\delta_R$ is given by:
\begin{eqnarray}
\delta_R \psi(x)&=&-igXt^a \psi(x)c^a(x) ,
\label{e-g}
\\
\delta_R A_\mu ^a(x)&=&{\Zt }\partial_\mu c^a(x) +
Xgc_{abc} c(x)^b A(x)_\mu^c =\Zt D_\mu c^a ,
\label{e-h}
\\
\delta_R c^a(x)&=&-{1\over 2}Xgc_{abc}c^b(x) c^c(x) ,
\label{e-i}
\\
\delta_R \bar c^a(x)&=&{1\over \xi}\,\partial\cdot  A^a(x) ,
\label{e-j}
\\
\delta_R h^\pm_v(x)&=&-igXt^a h^\pm_v(x)c^a(x) .
\label{e-k}
\end{eqnarray}
The full QCD lagrangian $\LQCD$ is also invariant
under~(\ref{e-g}-\ref{e-j}) and $\Psi$ transforming as $\psi$.
This well-known result will be used in section~\ref{sec4}.

Our goal is to prove that a clever choice of
$\ZYM$, $\ZH $, $\ZF$, $\Zt$, $X$ and $m_0$
suffices to render all Green functions of elementary fields,
as well as the composite
operators in~(\ref{e-g}--\ref{e-k}), and~(\ref{e2-a})
below, ultraviolet finite.

\section{\bf Renormalizability of QCD}
\label{sec2}

The proof of renormalizability that we sketch out in this section
requires that the regulated Green functions satisfy the Ward identities
implied by BRS invariance.
As already pointed out in the introduction, this is ensured by
using dimensional regularisation (or another regulator that does not break
BRS invariance).

Renormalizability of the standard QCD (light quark)
sector of the HQET with lagrangian $\Lqcd$ is proved by induction
on the number of loops, $N$.
The assumptions are the following:
\begin{enumerate}
\item All 1PI Green functions of elementary fields as well as those
with an insertion of the composite operators appearing in the BRS
transformations~(\ref{e-g}--\ref{e-j})
have successfully been made UV-finite at each order below $N$ loops.
This has been achieved by choosing the renormalization constants
$\ZF$, $\Zt$, $\ZYM$, $X$ and the bare mass $m_0$ appropriately.

\item As above for Green functions with insertions of the
operator
\beq
B_\mu^a(x)=\left({\Zt\over X}-1\right){1\over g} \partial_\mu \bar c(x)^a
+c_{abc} \bar c^b (x) A_\mu^c(x) .
\label{e2-a}
\eeq
This operator will be used to prove that $\de_R\psi$ (and
$\de_R h^\pm_v$, in section~\ref{sec3}),
and the quark-gluon vertex are UV-finite.
\end{enumerate}

Obviously~1 and~2 hold at tree level. It must be shown that they
also hold at order~$N$.
Thus, we proceed to compute $N$ loop contributions. Since subdivergences have
already been subtracted, only Green functions having non-negative superficial
degree of
divergence, $\delta(G)\ge 0$, may have an overall UV-divergence.
These
(potentially) UV-divergent Green functions are
the
quark, ghost and gluon self-energies; the ghost-gluon vertex; the gluon
three- and four-point functions; some (1PI) insertions of the operators
$\de_R A^a_\mu$,
$\de_R c^a$, $B^a_\mu$,
and $\de_R\psi$; and,
finally,
the  quark-gluon vertex.
Because of Weinberg's theorem, their
overall divergences are polynomials in the external momenta of
degree $\delta(G)$. Hence,
they may eventually
be absorbed in the counterterm lagrangian implied by~(\ref{e-c})
and the renormalization constants in~(\ref{e-g}--\ref{e-j}, \ref{e2-a}).

Next, {\bf i}) we choose $\ZF$ and $m_0$ to cancel the UV divergences of
the quark
two-point function; {\bf ii}) we choose $\ZYM$
to render the gluon self energy finite
(a BRS identity must be used to check that UV-divergences do not
show up in the longitudinal piece);
{\bf iii}) $\Zt $ is chosen to cancel the part of the divergence of
the ghost self-energy and {\bf iv}) $X$ is chosen to make
the ghost-gluon
vertex finite (note that only the colour tensor structure
$f_{abc}$ is allowed by charge
conjugation).
At this point, all renormalization constants have been
fixed and one must check that the UV-divergences of the remaining Green
functions automatically cancel against the relevant contributions
of the renormalization constants we have just determined.
This is accomplished through the use of BRS identities. One
proceeds orderly (for each step requires the conclusions of the preceding ones)
as follows:
{\bf v}) check that the operator $\de_R A^a_\mu$ is finite;
{\bf vi}) using ({\rm v}) and some BRS identity, check finiteness of
the
gluon three- and four-point functions; {\bf vii}) show that no UV-divergence
arises in insertions of $\de_R c^a$;
{\bf viii}) as above for $B^a_\mu$;
{\bf ix}) as above for $\de_R\psi$. Finally, use (i, ix) and
some identities to  prove that
{\bf x}) the quark-gluon vertex
is also UV-finite. This completes the proof of renormalizability.
(For details, see ref.\cite{c}.)

\section{\bf Renormalizability of the HQET}
\label{sec3}

Now, we turn to the HQET.
An important simplification arises from the observation that
loops of only heavy quark
propagators never occur.
This is due to the fact that in $\Lhqet$
heavy quarks, $h^+_v$,
$\bar h^+_v$, do not couple to
heavy antiquarks, $h^-_v$,
$\bar h^-_v$. Consequently,
the renormalization
of Green functions with no external effective heavy quark legs
(entirely described by $\Lqcd$)
remains
as in ordinary QCD.
Note that this statement
shows that steps ({\rm i}--{\rm x}) of the previous section can be carried out
exactly as
for the lagrangian~(\ref{e-c}) if no Green function with external heavy quark
fields is considered.
Among those Green functions involving $h^\pm_v$ there
are only four superficially divergent: the
quark two-point function, $\l h^\pm_v(x)\bar h^\pm_v(0)\r$; the
quark-gluon vertex, $\l h^\pm_v(x)\bar h^\pm_v(0) A^a_\mu(y)\r$,
as well as
$\l \de_R h^\pm_v(x) \bar h^\pm_v(0) \bar c^c(y)\r$ and
$\l  h^\pm_v(x) \de_R \bar h^\pm_v(0) \bar c^c(y)\r$. To the
induction assumptions of section~\ref{sec2}, one has to append the
finiteness of these
Green functions at each order below $N$ loops. One has also
to add to the previous list of steps the following: {\bf i'})
choose $\ZH$ to cancel
the UV-divergences in the heavy-quark two-point function;
{\bf ix'}) using the
appropriate identities, check that the operators $\de_R h^\pm_v$
and $\de_R \bar h^\pm_v$ are
finite; finally, {\bf x'}) show that no UV-divergence appears in the
(heavy) quark-gluon vertex. This, we carry out in this section.

Let us begin considering $\l h^\pm_v \bar h^\pm_v \r  $ in momentum space.
Weinberg's theorem tells us
that
the renormalization prescription can be chosen so that the overall
divergence of the 1PI portion of $\l h^\pm_v \bar h^\pm_v \r  $
is a polynomial in the external residual momenta or ``off-shellness''
{\em and} the light quark
mass of degree $\delta(G)$.
For such a renormalization prescription, the overall divergence can only be
proportional to $k\cdot v\, (1\pm\vs)/2$ and $m\, (1\pm\vs)/2$,
$k$ being the residual external
momentum and $m$ the mass of the
light quark propagating in the loops. Obviously, we can get
rid of the divergence proportional to $k\cdot v$
by adjusting $\ZH $. However, there is no counterterm in
$\LH$ to cancel a divergence proportional to the light
quark mass. Thus, we must show that this divergence is absent.
This is easily seen by noting that the $m$ dependence
of $\l h^\pm_v \bar h^\pm_v \r  $ comes only from light quark loops. The
corresponding traces of $\gamma$-matrices are seen to be an even
function of $m$
and,  hence, no UV divergence proportional to $m$ can arise.
This completes
step~({\rm i'}).

Now we turn to~({\rm ix'}). There is only one 1PI diagram with an insertion
of $\de_R h^\pm_v$ in which UV-divergences can occur:
\beq
\l \de_R h^\pm_v(x) \bar h^\pm_v(0) \bar c^c(y)\r^{1PI}.
\label{e3-a}
\eeq
To prove that~(\ref{e3-a}) is actually UV-finite,
we consider
$ \l \delta h^\pm_v(x)  \bar h^\pm_v(y) \Box \bar c^c(z)\r$. Clearly,
UV-finiteness
of this Green function ensures
UV-finiteness of~(\ref{e3-a}).
One can show that
\bea
\l \de_R h^\pm_v(x) \bar h^\pm_v(y) \Box	\bar c^a(z)\r  &=&
-{1\over X}\left\l \de_R h^\pm_v (x) \bar h^\pm_v (y)
{\delta S\over \delta c^a(z)}
\right\r
\nn\\
&-&
{1\over2}\xi g\l \de_R h^\pm_v (x) \;\de_R \bar h^\pm_v(y)
\;c_{ade}\bar c^d(z)\bar c^e(z)\r \label{e3-b}\\
&-&g\l \delta_R h^\pm_v (x)\; \bar h^\pm_v (y) \partial^\mu B_\mu^a(z) \r,\nn
\eea
where $S$ is the complete action, $S=\int d^Dx\,\Lhqet$. In order to
obtain~(\ref{e3-b}) we have computed $\partial^\mu B_\mu^a$
from~(\ref{e2-a}). Next, by~(\ref{e-j}) we have written $\partial^\mu A^a_\mu$
as $\xi \,\de_R\bar c^a$. Finally, we have used the nilpotence property
$\de_R [\de_R h^\pm_v]=0$ and the identity
$\de_R\l\de_R h^\pm_v\bar h^\pm_v \; c_{ade}\bar c^d\bar c^e\r=0$
to ``integrate by parts"
$\l\de_R h^\pm_v\bar h^\pm_v \, \de_R(c_{ade}\bar c^d\bar c^e)\r$
into its final form in~(\ref{e3-b}).

Each of the three terms of~(\ref{e3-b}) is now shown to be finite:
1) Using the anti-ghost equation of motion~\cite{c},
one can easily show
that
\begin{equation}
-{1\over X}\left\l \de_R h_v(x) \bar h_v (y) {\delta S\over \delta c^a(z)}
\right\r
=
g t^a\delta(x-z) \l h_v(x) \bar h_v(y)\r .
\label{e3-b_}
\end{equation}
Since the heavy quark two-point function is UV-finite by~({\rm i'}),
so is the first term of~(\ref{e3-b}).
2) The superficial degree of divergence of the second Green function
of~(\ref{e3-b})
is $\delta(G)=-1$. It is finite because it is 1PI and subdivergences
have already been canceled according to the induction assumptions.
3) Finally, we must consider the last term of~(\ref{e3-b}), whose decomposition
in 1PI parts is shown in figure~1.
For simplicity we have amputated the external quark leg,
which has previously seen to be UV-finite.
Graph~(c) has $\de(G)=-1$, so it is finite.
Adding graphs~(a) and~(b) one obtains
\beq
{\rm (a)}+{\rm (b)}=\l \de_R h^\pm_v \bar h^\pm_v
\bar c^b \r^{(N-1)}_{1PI} \l c^b \p^\mu B^a_\mu\r^{(N)} ,
\label{e3-b__}
\eeq
where the superscript $(L)$ indicates that contributions up to $L$ loops
are included. Note that the $N$-loop contribution from
$\l \de_R h^\pm_v \bar h^\pm_v
\bar c^b \r_{1PI}$ does not appear in~(\ref{e3-b__}) because no zeroth order
graph of $\l c^b \p^\mu B^a_\mu\r$ exists. The first factor
in~(\ref{e3-b__}) is finite by the induction assumption, while the second
factor is also finite
by step~({\rm viii}).
This completes
step~({\rm ix'}). Finiteness of $\de_R \bar h^\pm_v$ can be proved by the
same method.

Step~({\rm x'}) is an immediate consequence of~({\rm ix'}) and the
identity $\de_R\l h^\pm_v \bar h^\pm_v \bar c^a\r=0$,
which can be written as
\beq
{1\over\xi}\l h^\pm_v \bar h^\pm_v \partial^\mu A^a_\mu \r
=\l h^\pm_v \;\de_R \bar h^\pm_v \; \bar c^a\r-
\l \de_R h^\pm_v \; \bar h^\pm_v \; \bar c^a\r .
\label{e3-c}
\eeq
This Ward identity shows that the contraction of the
heavy quark gluon vertex (1PI, after subtraction of the UV-finite
external propagators)
with the momentum of the gluon leg has no UV divergence.
Since
by power counting the divergence of the vertex can only be proportional
to $v_\mu \,(1\pm\vs)/2$, the vertex itself is finite.
This completes the proof of renormalizability
of the HQET.

\section{Matching}
\label{sec4}

In the previous section we have shown that the
HQET
is renormalizable. As already mentioned,
this is precisely the main assumption used by Grinstein
in~\cite{g1}
to show that the Green functions of $\Lhqet$ and those of
$\LQCD=\LPsi+\Lqcd$
match for large $M$. Again, the proof in~\cite{g1} (and also here)
is by induction on $N$.
It is assumed that matching holds
at each order below $N$ loops for any Green function of the type considered
in the induction assumptions~1 and~2 of section~\ref{sec2}.
Note that at tree-level
any Green function matches by construction~\cite{georgi}.
Then, as
shown in~\cite{g1}, the
matching at order $N$ holds for graphs with negative superficial degree of
divergence.
Thus, we need only be concerned with the matching of
$\l\Psi\bar\Psi\r$, $\l\Psi\bar\Psi A^a_\mu\r$,
$\l\de_R\Psi\bar\Psi \bar c^a \r$
and $\l\Psi \de_R \bar\Psi\bar c^a \r$.

Let us begin by borrowing from~\cite{g1} the following result:
it is always possible to modify the counterterm in
$\ZPsi$, $\ZH$ and $M_0$ by an UV-finite amount
in such a way that the quark two-point functions match,
i.e, eq.~(\ref{e1-a})
is satisfied, at $N$ loops. One can check that the readjusted
renormalization constants
and $M_0$ become
functions of $\log M$. If BRS invariance is to be preserved, one cannot
change the value of the remaining counterterm in~(\ref{e-_g}) (see
also~eq.~(\ref{e-e})). Therefore,
we must show that the QCD and QHET quark-gluon vertexes match automatically.

First, we shall check that
\beq
\l \de_R \Psi \; \bar \Psi \; \bar c^a\r|_{\pm Mv+p}
-\l \de_R h^\pm_v \; \bar h^\pm_v \; \bar c^a\r|_p=O\left({1\over M}\right) .
\label{e4-e}
\eeq
Since
the identities obtained from~~(\ref{e3-b}), (\ref{e3-b_}) and~(\ref{e3-b__})
by the replacement $h^\pm_v \rightarrow \Psi$ also hold, we only need
to prove the matching of the
Green functions on the right hand side
of~(\ref{e3-b}) with their $\Psi$~counterparts (in momentum space).
The matching of
$\l\de_R h^\pm_v\bar h^\pm_v \de S/\de c^a\r$
and $\l\de_R \Psi\bar \Psi \de S/\de c^a\r$ is guaranteed
by
the matching of the quark two-point functions~(\ref{e1-a}) and
by the $\Psi$- and $h_v$-version of~(\ref{e3-b_}), which in momentum space
read
\beq
-{1\over X}\left\l \de_R \chi\bar\chi{\delta S\over \delta c^a}
\right\r
=
g t^a \l \chi \bar \chi\r,\qquad \chi=h^\pm_v,\Psi.
\eeq
The matching of $\l\de_R h^\pm_v \de_R \bar h^\pm_v
\,c_{ade}\bar c^d\bar c^e\r$ and
$\l\de_R \Psi \de_R \bar \Psi
\,c_{ade}\bar c^d\bar c^e\r$ holds because both have negative superficial
degree of
divergence. Finally, we must prove that $\l\de_R h^\pm_v \bar h^\pm_v
B^a_\mu \r$ and $\l \de_R \Psi \bar \Psi
B^a_\mu \r$ also match. This is apparent from figure~1
since
graphs of the type~(c) have $\de(G)=-1$, whereas from~(\ref{e3-b__}),
graphs~(a) and~(b) are seen to involve
$\l \de_R h^\pm_v \bar h^\pm_v
\bar c^b \r_{1PI}$ and
$\l \de_R \Psi \bar\Psi
\bar c^b \r_{1PI}$
at $N-1$ loops {\em at most}, so they match by the induction assumption.
It is important to recall that
the external quark legs have been
amputated in figure~1. This causes
no trouble, for the quark propagators
are already known to match.
Similarly, one can show that the matching of the terms involving
$\de_R \bar\Psi$ and $\de_R \bar h^\pm_v$
also holds.

We now turn to the quark-gluon vertex. One can show~\cite{g1} that
\beq
\l\Psi\bar\Psi A^a_\mu\r|_{\pm Mv+p}-
\l h^\pm_v\bar h^\pm_v A^a_\mu\r|_{p}=
   \Pv t^a \Cc^\pm v_\mu \Pv+O\left({1\over M}\right) .
\label{e4-c}
\eeq
Where $\Pv =(1\pm \vs)/\; 2p\cdot v$
is the effective heavy quark propagator and the coefficient
$\Cc^\pm$ is UV-finite, dimensionless and dependent only on $\log M$.
For simplicity we have amputated the gluon leg in~(\ref{e4-c}).
It must be shown that $\Cc^\pm$ vanishes.
This is easily seen from~(\ref{e3-c}),
the Ward identity
\beq
{1\over\xi}\l \Psi \bar \Psi \partial^\mu A^a_\mu \r
=\l \Psi \;\de_R \bar \Psi \; \bar c^a\r-
\l \de_R \Psi \; \bar \Psi \; \bar c^a\r
\label{e4-d}
\eeq
and the matching of the Green functions
on their right hand sides. This completes the proof.

\newpage

\section{Conclusions and discussion}
\label{sec5}

We have shown that the HQET is renormalizable
perturbatively. Weinberg's theorem and BRS invariance
play an essential role in the
proof. Weinberg's theorem is assumed to hold
for the HQET, as several one- and two-loop calculations seem
to indicate,
while BRS invariance is explicitly
preserved at any order
in perturbation theory.
We have also shown that the HQET provides the
infinite mass limit
of QCD for heavy quarks near mass-shell. Again,
BRS
invariance is crucial in our proof.
If the \MSb\ scheme is chosen, QCD and the HQET match in
the following
sense:
$\l \Psi \bar\Psi\; \cdots\r|^{\overline{\rm MS}}_{\pm Mv+p}
=C(\log M)\l h^\pm_v \bar h^\pm_v \;\cdots
\r|^{\overline{\rm MS}}_p+O(1/M)$,
where the dots stand for
any string of (light) fields and $C^{1/2}(\log M)$ is a
{\em UV-finite}
wave function renormalization constant of the heavy quark fields.

\bigskip

\noindent {\bf Acknowledgments} We thank P.~Ball for
a fruitful discussion. We are grateful to J.~Soto and
R.~Tarrach for reading the manuscript. P.~G. acknowledges
gratefully a grant from
the {\em Generalitat de Catalunya}.

\newpage

\begin{flushleft}
{\large\bf Figure captions}
\end{flushleft}

\begin{description}

\item[Fig.~1.] Decomposition of
$\l \delta_R h^\pm_v (x)\; \bar h^\pm_v (y) \partial^\mu B_\mu^a(z) \r$
in 1PI portions. For simplicity the external quark leg has been amputated.
The Feynman rules for the two different insertions of
$\partial^\mu B_\mu^a$ can be read off from the corresponding two
terms of~(\protect\ref{e2-a}).
The same decomposition holds in QCD by substituting~$\Psi$
for~$h^\pm_v$.

\end{description}

\newpage

\end{document}